# All-Optical Nonlinear Pre-Compensation of Long-Reach Unrepeatered Systems

Pawel M. Kaminski[(1)]*, Tiago Sutili[(2)], José Hélio da Cruz Júnior[(2)], Glauco C. C. P. Simões[(2)], Francesco Da Ros[(1)], Metodi P. Yankov[(1)], Henrik E. Hansen[(1)], Anders T. Clausen[(1)], Søren Forchhammer[(1)], Leif K. Oxenløwe[(1)], Rafael C. Figueiredo[(2)], and Michael Galili[(1)]

[(1)] DTU Fotonik, Technical University of Denmark, DK-2800 Lyngby, Denmark, pkam@fotonik.dtu.dk
[(2)] Optical Communication Solutions, CPQD, BR-13086-902 Campinas, Brazil

**Abstract** *We numerically demonstrate an all-optical nonlinearity pre-compensation module for state-of-the-art long-reach Raman-amplified unrepeatered links. The compensator design is optimized in terms of propagation symmetry to maximize the performance gains under WDM transmission, achieving 4.0dB and 2.6dB of SNR improvement for 250-km and 350-km links.*

## Introduction

Optical technologies are currently being pushed to their limits due to the ever-increasing traffic demands[1]. In particular, the inherent nonlinearity of silica glass is causing difficult-to-compensate signal degradations in fiber-optic systems. These Kerr-related impairments currently set the upper-bound on achievable information rates and transmission distances[1]. Consequently, there has been significant effort to compensate nonlinear degradations, and it is generally possible in both digital and optical domains. Digital methods often require substantial processing time, and are unsuitable for multi-channel applications due to limited receiver bandwidth[1]-[3]. On the other hand, optical techniques are not only instantaneous, but also capable of simultaneous processing of all frequency channels in wavelength-division-multiplexed (WDM) systems. Among them, optical-phase conjugation (OPC) has been shown to successfully address nonlinear distortions, but it heavily relies on specific link design to achieve a relevant performance improvement[4],[5]. In particular, symmetric pulse propagation is required on each side of the OPC device, which is challenging to achieve in most practical applications. Moreover, the standard OPC approach is unsuitable for unrepeatered transmission, where it is not feasible to perform optical conjugation along the link. For such systems, lumped compensation modules based on OPC have been previously considered at both the transmitter[6] and receiver[7],[8]. However, they were limited to standard links using erbium-doped fiber amplifiers (EDFA), and achieved only partial propagation symmetry due to the power profile mismatch between the compensator and the link. A numerical analysis from[9] was able to show strong system symmetry, but relied on unrealistic fiber types for compensation, and thus it remains unattainable in practice.

In this work, we revisit the OPC requirements, and numerically demonstrate how the technique can be applied to a state-of-the-art Raman-amplified unrepeatered link by using an OPC-based pre-compensation module at the transmitter side. We only consider commercially available equipment and optimize the symmetry of propagation through the design of the optical pre-compensation unit by means of: (1) scaled down compensation medium, (2) novel dispersion mapping, and (3) Raman amplification for fine power profile tuning. Ultimately, we establish a high degree of propagation symmetry, and achieve net signal-to-noise ratio (SNR) improvements ranging from 2.6 dB up to 4.0 dB, depending on the link configuration. The standard split-step Fourier method (SSFM) is employed for the propagation, and the particle swarm optimization (PSO) algorithm is used for finding the optimum system parameters.

## Symmetry Requirements

The fundamental principle of OPC is based on reversing the sign of distortions by phase-conjugation of the optical field, and reapplying them throughout the transmission system. Perfect cancellation requires identical accumulated nonlinear phase-shifts to be induced on both sides of the OPC device, for the signal and its conjugated copy, respectively. As the phase-shift is both power and pulse shape dependent, it is necessary to continuously restore the nominal power at the correct pulse shapes over the entire propagation distance. Pulse shape is predominantly determined by the accumulated dispersion, and so it is common to express the degree of nonlinearity matching through power versus accumulated dispersion diagrams (PADD), where perfect compensation is attained for identical powers at the opposite values of accumulated dispersion[10],[11]. The diagrams accurately predict the symmetry for homogenous links, but they ignore a possible mismatch of other fiber parameters. Instead of focusing on power alone, in this work we express the nonlinear phase-shift as $\Phi_{NL} = \int \frac{\gamma P}{D} dD_{acc}$ by using $D_{acc} = L * D$, and employ the integral to completely describe the symmetry through more general nonlinearity versus accumulated dispersion diagrams (NADD). The advantage of this scheme over PADD is illustrated in Fig. 1. In this example (Fig. 1a), the length of the second span is only half of the length of the first span, while the power and dispersion in the second span are doubled to induce the same pulse evolution over a smaller distance. As the powers of the signal (blue) and the conjugate (red) are not strictly matched, the PADD of this link is highly asymmetric (Fig. 1b). However, the corresponding NADD (Fig. 1c) accurately depicts the matching of nonlinearities because it also incorporates the other propagation parameters.

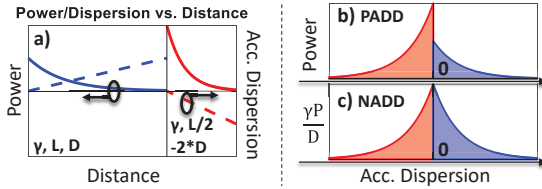

**Fig. 1:** a) Symmetric system with unequal powers and lengths, with the corresponding PADD (b) and NADD (c). Propagation of signal (blue) and conjugate (red).

It is also noted that we employ a nonstandard power/dispersion symmetry from[12],[13], where each span accumulates dispersion with opposite sign, both starting from zero. This simplifies the design, but requires a lumped dispersion compensating module in between the spans to compensate the dispersion of the first part.

**Transmission Link**
We aim to apply the above OPC design principles to a state-of-the art unrepeatered link, shown in Fig. 2. Record transmission performance has been recently reported with such systems[3],[14],[15]. It is composed of 100 km of large effective area fiber (Vascade EX2000), followed by 250 km of low-loss single mode fiber (LL-SMF). Additionally, it is supplemented with bi-directional Raman amplification (two high-power Raman pumps on both sides, each pump providing up to 23 dBm power), and a remote optically-pumped amplifier (ROPA) at 250 km. The power distribution along the link is presented alongside the system in Fig. 2. Observe that the nonlinearity is predominantly generated at the beginning of the link. This part is therefore specifically targeted with the pre-compensation setup. We conduct the analysis for two separate scenarios: 1) the forward Raman-pumped link up to 250 km (blue), and 2) the full 350 km system including ROPA and backward Raman-pumping (blue and black). Both scenarios use the same pre-compensation scheme, but it is finely tuned in each case to maximize the performance.

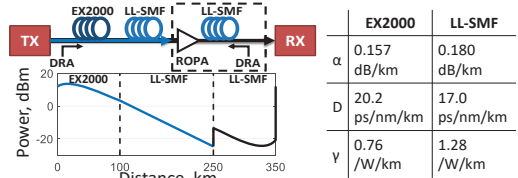

**Fig. 2:** Sketch of the unrepeatered transmission system under analysis and the power evolution: (blue) forward-pumped 250 km link, and (black) backward-pumped 100 km link extension.

**Pre-Compensated System**
The proposed pre-compensation setup is presented in Fig. 3, jointly with the rest of the simulation system. At the transmitter side, seven WDM channels are generated on a 37.5 GHz grid with carrier linewidth ν=10 kHz. They are subsequently modulated with 16-QAM data at 32 GBd per polarization, combined altogether and amplified. Noise figures of all EDFA's in the system are set to NF=5 dB. The signal then enters the optical pre-compensation unit. The appropriate nonlinear phase-shift is first induced in a dispersion compensating fiber (DCF: $\alpha_0$=0.50 dB/km; $D_0$=-100.0 ps/nm/km; $\gamma_0$=5.40/W/km) with forward Raman-pumping to allow for fine-tuning of the power profile shape. The Raman noise is added uniformly across the entire frequency band. DCF is chosen as the nonlinear medium because it is much more dispersive than the link, and so it allows for maximum down-scaling of the compensation unit length. After the DCF, the accumulated dispersion is fully compensated using a lumped dispersion compensating module (e.g. a fiber Bragg-grating (FBG)) to ensure pulse shape matching before and after the OPC. The phase-shift is then reversed using an ideal OPC, which conjugates the complex field amplitude $A(z,t) \rightarrow A^*(z,t)$ with no penalties associated with the process. Such pre-distorted signal enters the unrepeatered link for transmission. At the end of the link, it is received with a standard coherent receiver followed by digital signal processing (DSP) chain. The SNR of the central channel is estimated from the transmitted and received symbols[16], as $SNR = \mathbb{E}_k[|x_k|^2]/\mathbb{E}_k[|y_k - x_k|^2]$.

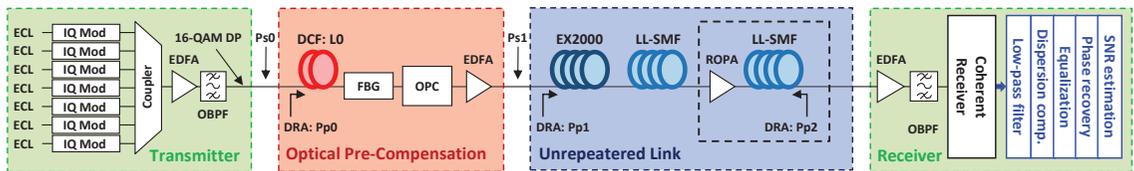

**Fig. 3:** Complete simulation setup for the system optimization and evaluation of performance gains.

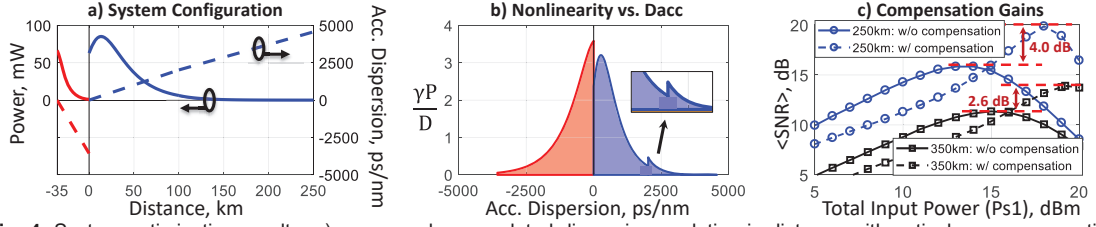

**Fig. 4:** System optimization results: a) power and accumulated dispersion evolution in distance with optical pre-compensation included (for the 250 km link), b) nonlinearity vs. accumulated dispersion diagram outlining system symmetry (for the 250 km link), and c) net relative gains in SNR for the 250 km link (blue), and the full 350 km system (black).

PSO algorithm is employed to maximize the SNR by joint optimization of the compensation unit and the link (see Fig. 3) in terms of:
- signal launch power into DCF (Ps0);
- Raman pump power into DCF (Pp0);
- DCF compensation length (L0);
- signal launch power into the link (Ps1);
- Raman pump power into the link (Pp1).

The PSO optimization (acceleration coefficients: C1 = 1.2 and C2=0.1; inertia W=0.05) is carried out until convergence using 100 randomly initialized particles[17]. The parameters are optimized for two scenarios: 250 km link and the full system including the final part (marked with dashed lines in Fig. 3). For the latter, backward Raman pumping (Pp2) is fixed at 23 dBm, and ROPA operates at a constant 10 dB gain.

**Simulation Results**

The optimized parameters for both scenarios including optical pre-compensation are presented in Table 1, and the resultant performance improvement is illustrated in Fig. 4.

**Table 1:** PSO results for 250 km (Link1) and 350 km (Link2).

|  | L0 [km] | Ps0 [dBm] | Pp0 [dBm] | Ps1 [dBm] | Pp1 [dBm] |
|---|---|---|---|---|---|
| Link1 | 35.5 | 18.2 | 16.0 | 18.0 | 23.0 |
| Link2 | 29.0 | 19.2 | 16.8 | 19.2 | 23.0 |

The key difference between the links is a boost for Ps0 and Ps1, as transmission distance increases. The power and dispersion evolution in distance are shown in Fig. 4a for the 250 km scenario. In Fig 4b we illustrate the nonlinearity versus accumulated dispersion of this system, revealing a high degree of symmetry. The discontinuity at accumulated dispersion of 2020 ps/nm is caused by a change of transmission fiber (from EX2000 to LL-SMF). SNR gains for both link configurations are presented in Fig. 4c. For all cases, the parameters are fixed to the optimum, but only the launch power Ps1 is varied to illustrate the matching of nonlinearities. Without optical pre-compensation, the forward Raman-pump is operating at the optimum Pp1=23 dBm. The simulated performance for the uncompensated system is consistent with our experimental analysis of a similar 350 km unrepeatered link[3], where 11.7 dB of SNR was attained without nonlinearity compensation, and it was increased to 12.0 dB by digital back-propagation. Note that the results in [3] are given in terms of mutual information. In this analysis we observe relative SNR penalty from inclusion of the compensator when the input power deviates from the optimum (Fig. 4c), and the nonlinearity is not matched. However, with the launch power tuned properly, the nonlinear phase-shifts from the DCF and the link are complementary, leading up to 4.0 dB and 2.6 dB net gains in SNR for the shorter link, and for the full system, respectively. The lower net improvement for the full system is mostly due to an increased contribution of the amplified spontaneous emission noise from the low-power link extension. Overall, the proposed pre-compensator shows great potential for improving unrepeatered systems' performance.

**Conclusions**

In this work, we proposed a general method for achieving OPC symmetry through more accurate nonlinearity versus accumulated dispersion mapping. We applied this technique to a state-of-the-art Raman-amplified unrepeatered system, and we designed and optimized an all-optical nonlinearity pre-compensation module for it. The unit consists of three main components: DCF for nonlinear pre-distortion, FBG for dispersion matching, and OPC for phase-shift reversal. We were able to engineer the symmetry of the system, scale down the compensation structure to a fraction of the link, and provide efficient nonlinearity suppression for seven channel WDM transmission. PSO algorithm was employed to tune the setup parameters, leading to 4.0 dB gain in SNR for a 250 km link, and 2.6 dB for a 350 km system including ROPA and backward pumping.

**Acknowledgments**

The work was supported by the DNRF Research CoE, SPOC (ref. DNRF123), the ERC CoG FRECOM (771878), the Villum YIP OPTIC-AI (29344), and by the Brazilian MCTI, FUNTTEL/Finep (01.19.0088.00), and CNPq.